\documentclass[a4paper]{jpconf}
\usepackage{a4wide,amsmath,amssymb,alltt,graphicx,enumitem}

\newcommand\ie{i.e.\ }
\newcommand\eg{e.g.\ }

\newcommand\lbrac{\symbol{123}}
\newcommand\rbrac{\symbol{125}}
\newcommand\brac[1]{\lbrac#1\rbrac}
\newcommand\uscore{\symbol{95}}
\newcommand\Code[1]{\ensuremath{\texttt{#1}}}
\newcommand\Var[1]{\ensuremath{\mathit{#1}}}
\newcommand\button[1]{\lower .45ex\hbox{\includegraphics[height=2.2ex]{button_#1}}}

\renewcommand{\thefootnote}{\fnsymbol{footnote}}

\begin{document}


\title{New Features in FeynArts \& Friends, \\
and how they got used in FeynHiggs}

\author{T.~Hahn$^1$}

\address{${}^1$
  Max-Planck-Institut f\"ur Physik,
  F\"ohringer Ring 6, D--80805 Munich}

\date{\today}

\begin{abstract}
\noindent
This note gives an update on recent developments in FeynArts, FormCalc,
and LoopTools, and shows how the new features were used in making the 
latest version of FeynHiggs.
\hfill MPP--2019--112
\end{abstract}


\section{Introduction}

The FeynArts \cite{FeynArts}, FormCalc \& LoopTools \cite{FormCalc} 
triad of packages are used for the generation and calculation of Feynman 
diagrams up to one loop.  With QCD and Standard Model calculations 
moving to higher orders their main focus has shifted somewhat to BSM 
models and package building, where the availability of analytical 
results in conjunction with a toolkit of functions and options in 
Mathematica, rather than a monolithic program, has proven very useful in 
implementing non-standard tasks like special approximations, extraction 
of coefficients, or nontrivial renormalizations \cite{HahnPassehr}.

FormCalc's code generator has been used a lot for package building,
for FeynHiggs \cite{FeynHiggs} as described in more detail in the
following but also \eg in SARAH \cite{SARAH} and SloopS \cite{SloopS}.

The remainder of this note starts with a overview of the code structure 
of FeynHiggs, discusses improvements made to the generated parts, and 
then describes the new functions and options which were added to 
FeynArts and FormCalc to achieve this.


\section{FeynHiggs Code Structure}

FeynHiggs est omnis divisa in partes tres.
\begin{enumerate}
\item Code hand-written for FeynHiggs

The `back bone' of FeynHiggs is of course written by hand.  This includes
all structural code, utility functions, and the contributions taken from
literature.

\item Code generated from external expressions

In several cases analytical expressions are available from external
sources, usually in Mathematica format, from which parts of the FeynHiggs
code are generated.  Examples are the two-loop Higgs self-energies, some 
EFT ingredients, the muon $g_\mu{-}2$, or the two-loop parts of $\Delta r$.

This is an improvement over hand-coded expressions since small changes
can easily be applied, and the code can be optimized and re-generated as 
necessary.

\item Code generated from calculations done in/for FeynHiggs

Full control over model content, particle selection, renormalization 
prescription, resum\-mations, etc.\ is achieved in the calculations done 
for and included in FeynHiggs, in the \Code{gen} subdirectory.  
Complete, model-independent automatization, sometimes dubbed the 
`generator generator' approach, is not intended as the produced code 
still needs to be embedded in and called from the main program, in which 
the inputs have to be properly adjusted.

\end{enumerate}


\section{Improvements in FeynHiggs Code Generation}

The unrenormalized one-loop Higgs self-energies have been generated with a
high degree of automatization since FeynHiggs 2.0.  Before version 2.14,
however, the entire renormalization was hard-coded.

The new procedure instead reads the renormalization (counter-terms plus 
renormalization constants) from the model file \cite{MSSMCT}.  It knows 
about a few flags governing the renormalization, such as 
\Code{\$MHpInput}, which selects whether $M_A$ or $M_{H^+}$ is the input 
mass for the Higgs sector, but otherwise makes as few model assumptions 
as possible. All renormalized self-energies are split into parts 
($t/\tilde t$; $t/\tilde t$+$b/\tilde b$; all $f/\tilde f$; all) to 
preserve the possibility to look at individual sectors only, though this 
is mostly done for comparison and debugging today.

To achieve the desired level of automation, several of FormCalc's
code-generation functions had to be enhanced or added.


\section{Enhanced and New FormCalc Functions}

\subsection{Declarations and Code in One File}

A generated file containing both declarations and code cannot 
straightforwardly be included in a language like Fortran, which requires 
the strict order ``declarations before code.''

This is addressed through the new $\Code{DeclIf}\to\Var{var}$ option of 
FormCalc's \Code{WriteExpr}, which inserts preprocessor statements as 
follows:
\begin{alltt}
   #ifndef \Var{var}
   #define \Var{var}
     (declarations)
   #else
     (code)
   #endif
\end{alltt}
The \Var{var} argument must be a variable name acceptable to the
preprocessor but has no purpose beyond that.  The file thus created can 
be included twice, once in the declarations section and once in the 
code section.


\subsection{Temporary Variables}

FormCalc's code generator has already in the past had the notion of 
`temporary variables' which would typically be introduced through 
optimization, \ie when duplicate code was detected it was computed 
once and stored in a temporary variable.

Temporary variables can now also be added by the user explicitly, 
through the \Code{MakeTmp} option of \Code{PrepareExpr}.  Its argument
is a function which is applied to the input expressions and may insert 
any number of variable definitions ($\Var{var}\to\Var{value}$) into 
the list.

Usually the new function 
\Code{ToVars[\Var{patt},\,\Var{name}][\Var{exprlist}]} is used with 
\Code{MakeTmp}, which puts subexpressions that match the pattern 
\Var{patt} into variables that begin with \Var{name}.  Example:
\begin{alltt}
   WriteExpr[\Var{expr}, MakeTmp \(\to\) ToVars[LoopIntegrals,\,Head]]
\end{alltt}
The second argument \Code{Head} means to take the head of the
abbreviated expression for the variable name, which in the case of loop 
integrals might result in names like \Code{C0i7}, \Code{D0i3}, etc.

The advantages are twofold:
Firstly the effect of optimization can be enhanced, say if
a loop integral that does not depend on an index is abbreviated
from a larger expression that does, then the loop integral can be
hoisted outside the loop over the index.

Secondly, the abbreviated variables can be inspected at run-time
using the debugging apparatus for generated code (\Code{DebugLines}, 
\Code{\$DebugCmd}).


\subsection{Improved Abbreviations}

Also the abbreviations introduced through \Code{Abbreviate} have been 
improved.  The function \Code{Abbreviate} still works in two modes:
\begin{alltt}
   Abbreviate[\Var{expr},\,\Var{level}]
   Abbreviate[\Var{expr},\,\Var{func}]
\end{alltt}
where the \Var{level} (integer) specification [unchanged] means to 
introduce abbreviations below that level in the expression tree.

The \Var{func} (function) mode also traverses the expression tree but 
for every subexpression `asks' \Var{func} whether an abbreviation should 
be introduced.  Allowed responses were \Code{True} and \Code{False} so 
far; it can now also be an expression, usually a part of its argument, 
for which the abbreviation will then be introduced.  For example, a 
numerical prefactor can be stripped off to avoid having many 
abbreviations that differ only by a number.

Secondly, the list of abbreviations returned with \Code{Abbr} and
\Code{Subexpr} now has proper patterns on the left-hand sides.  
Outwardly the effect seems rather minuscule, \eg
\begin{alltt}
(old) Sub333[Gen5] \(\to\) A0[Mf2[2,Gen5]] - ...
(new) Sub333[Gen5_] \(\to\) A0[Mf2[2,Gen5]] - ...
\end{alltt}
but while the old version was in general good only for write-out to 
a Fortran program, the new version can actually be used to substitute
Mathematica expressions.


\subsection{Finding Dependencies}

The new FormCalc function \Code{FindDeps[\Var{list},\,\Var{patt}]} finds 
all variables in \Var{list} whose r.h.s.\ directly or indirectly depends 
on \Var{patt}.

Example:
\Code{FindDeps[\brac{a\,$\to$\,x, b\,$\to$\,2, c\,$\to$\,3\,+\,a, 
d\,$\to$\,b\,+\,c}, x]} gives \Code{\brac{a, c, d}}.


\subsection{Named Array Indices}

A special treatment of named array indices may not seem necessary
since Mathematica can itself deal with `indices' of any type.  For 
the automated generation of declarations the array dimensions must be
known, however, and they cannot be inferred from named indices.

The new FormCalc function \Code{Enum} associates labels with numbers 
and works similarly to the C \Code{enum} keyword.  
\Code{ClearEnum} removes all such associations again.
Label definitions are used for dimension computations only, never
actually substituted in expressions.

Example: \Code{Enum["h0h0", "HHHH"\,$\to$\,3, "h0HH"]} respectively 
associates the strings \Code{h0h0}, \Code{HHHH}, and \Code{h0HH} with 
array indices 1, 3, and 4.


\subsection{Persistent Names for Generic Objects}

FeynArts' Generic amplitudes contain objects not representable in FORM 
(or later on in Fortran), such as 
\Code{G[\Var{s}][\Var{cto}][\Var{fi}][\Var{kin}]} for a generic 
coupling.  \Code{CalcFeynAmp} must replace them by simple names before 
the computation can commence.

In former versions numbered symbols were used, \eg \Code{Coupling5}, 
which were of course not persistent beyond one FormCalc session.  
This was never a problem with computations at deeper levels, where 
the generic identifiers got substituted by their actual values even 
before \Code{CalcFeynAmp} returned.

To make the Generic amplitude useful by itself a portable name-mangling 
scheme has now been implemented.  This allows to produce Generic 
`building blocks' for applications, though at the price of rather 
unwieldy symbol names like \Code{GV1VbtVbbg12Kp3g23Pq1g13kQ2}.


\subsection{Propagator-dependent Masses and Vertices}

FeynArts can now distinguish masses and couplings for different 
propagator types.  A particle can have different masses inside and outside
of a loop, for example.  The change in syntax is such that existing 
model files are not affected.

In the particle list \Code{M\$ClassesDescription} a loop-level mass 
would be declared as \eg
\begin{alltt}
   S[1] == \brac{..., Mass\:\(\to\)\:MHtree, Mass[Loop]\:\(\to\)\:MH, ...}
\end{alltt}
The couplings in \Code{M\$CouplingMatrices} can similarly be extended as 
\eg
\begin{alltt}
   C[S[1,t1], S[2,t2], S[2,t3]] == \Var{coupling}
\end{alltt}
except that in this case the \Code{t1,2,3} on the left must be
placeholders (symbols), not literals as `\Code{Loop}', and may appear
in the \Var{coupling} function on the r.h.s., \eg in an 
\Code{If}-statement.


\subsection{Changes for Mixing Fields}

Mixing fields propagate as themselves but couple as their left and
right partners.  A classic example is the $G^0$--$Z$ and
$G^\pm$--$W^\pm$ mixing in the Standard Model in a non-Feynman gauge.

Reversed mixing fields used to be represented in FeynArts as
\Code{Rev[\Var{g},\Var{g'}]} at Generic level, but as
\Code{2~Mix[\Var{g},\Var{g'}]} at Classes level.
This lead to inconsistencies (too many/few diagrams) so that now the
reversed field is represented by \Code{Rev[\Var{g},\Var{g'}]}
also at Classes level.

Unfortunately this change does affect existing model files, though mixing 
fields are not a very commonly implemented feature.


\section{Mixed Precision in One Code}

The numerical stability of FeynHiggs is generally satisfactory but \eg
the non-degenerate two-loop EFT threshold corrections exhibit numerical
artifacts even in not-too-extreme scenarios.

All-out quadruple precision has been available for long
(\verb=./configure --quad=) but is much slower than double precision.
For everyday use higher precision must be restricted to critical parts.
A simple solution portable across all major Fortran compilers 
turned out to be a `poor man's template programming' using the 
preprocessor.  The relevant part of \Code{types.h} reads

\begin{verbatim}
#ifdef QuadPrec
#  define RealSize 16
#  define ComplexSize 32
#  define RealSuffix Q
#else
#  define RealSize 8
#  define ComplexSize 16
#  define RealSuffix D
#endif
#define RealQuad real*RealSize
#define ComplexQuad complex*ComplexSize
#define _id(s) s
#define _R(s) _id(s)RealSuffix
#define N(n) _id(n)_id(_)RealSize
\end{verbatim}
Three aspects are addressed here:
\begin{itemize}
\item extended real and complex types (\Code{Real,ComplexQuad}),
\item treatment of number literals (\Code{N(1.234)}), and
\item name mangling for routines needed in more than one precision 
(\Code{\uscore R(routine)}).
\end{itemize}
Code outfitted with these macros can be switched from double to quadruple 
precision by just defining \Code{QuadPrec}.  The increase in the overall 
runtime due to the extra precision required by a handful of routines was 
in the end hardly noticeable.


\section{Evaluation of generic Mathematica Expressions in FORM}

Sending Mathematica expressions to FORM for fast evaluation is one of 
the central principles of FormCalc, hence the name.  Beyond the speed 
aspect there is also a curious complementarity of instruction sets which 
makes some operations vastly simpler in FORM (and conversely others in 
Mathematica), for example removing terms higher than a certain power in 
an expansion is done in FORM with a mere declaration.

FormCalc's interfacing code is now available in a separate package, 
FormRun, available from the FormCalc Web page.  It implements a function 
of the same name used as
\begin{alltt}
   FormRun[\Var{exprlist},\,\Var{decl},\,\Var{cmd}]
\end{alltt}
This sends \Var{exprlist} to FORM for evaluation, with \Var{decl} 
extra declarations and \Var{cmd} FORM commands to be executed.  Both 
\Var{decl} and \Var{cmd} are optional.  Variables for which no specific 
declaration is given are symbols in FORM.  There is no protection 
against expressions FORM cannot represent (\eg \Code{h[1][2]}).

The input \Var{exprlist} can be just a single, unnamed expression, but 
the output will always be a list of expressions, simply because the 
FORM output may contain more than one expression.  Vector and matrix 
expressions are returned element-wise, \eg
\Code{M}\,$\to$\,\Var{(matrix)} comes back as a list of expressions 
\Code{M11}, \Code{M12}, etc.  An unnamed expression comes back as 
\Code{expr}.


\section{Summary}

This note describes many small functions and additions to FeynArts, 
FormCalc \& LoopTools, mostly triggered by FeynHiggs development.
Together they yield significant improvements, in particular in code 
generation.

\bigskip



\begin{thebibliography}{99}

\raggedright

\newcommand{\volyearpage}[3]{\textbf{#1} (#2) #3}
\newcommand{\cpc}{\textsl{Comp.\ Phys.\ Commun.} \volyearpage}
\newcommand{\jpc}{\textsl{J.\ Comp.\ Phys.} \volyearpage}
\newcommand{\cip}{\textsl{Comp.\ in Phys.} \volyearpage}
\newcommand{\toms}{\textsl{ACM Trans.\ Math.\ Software} \volyearpage}
\newcommand{\tomacs}{\textsl{ACM Trans.\ Modeling Comp.\ Simulation} \volyearpage}
\newcommand{\siam}{\textsl{SIAM J.\ Numer.\ Anal.} \volyearpage}
\newcommand{\numa}{\textsl{Numer.\ Math.} \volyearpage}

\bibitem{FeynArts}
  T.~Hahn,
  Comput. Phys. Commun. \textbf{140} (2001) 418 [hep-ph/0012260].

\bibitem{FormCalc}
  T.~Hahn, M.~Perez-Victoria,
  Comput. Phys. Commun. \textbf{118} (1999) 153 [hep-ph/9807565].

\bibitem{HahnPassehr}
  T.~Hahn, S.~Pa\ss ehr,
  Comput. Phys. Commun. \textbf{214} (2017) 91 [arXiv:1508.00562].

\bibitem{FeynHiggs}
  M.~Frank et al.
  JHEP \textbf{0702} (2007) 047 [hep-ph/0611326];
  H.~Bahl et al.
  arXiv:1811.09073.

\bibitem{SARAH}
  F.~Staub,
  Comp. Phys. Commun. \textbf{185} (2014) 1773 [arXiv:1309.7223].

\bibitem{SloopS}
  N.~Baro, F.~Boudjema, A.~Semenov,
  Phys. Rev. \textbf{D78} (2008) 115003 [arXiv:0807.4668].

\bibitem{MSSMCT}
  T.~Fritzsche et al.
  Comp. Phys. Commun. \textbf{185} (2014) 1529 [arXiv:1309.1692].

\end{thebibliography}
\end{document}